\documentclass[aps,prd,twocolumn,amsmath,superscriptaddress,amssymb,showpacs,floatfix,nofootinbib,longbibliography]{revtex4-1}
\pdfoutput=1
\usepackage{graphicx}
\usepackage{placeins}
\usepackage{bm}
\usepackage{times}
\usepackage{slashed}
\usepackage{color}
\usepackage{aas_macros}

\usepackage{slashed}
\usepackage{lipsum}
\usepackage{subfigure}
\usepackage{multirow}
\usepackage{amsmath}
\usepackage{array} 
\usepackage{varwidth} 
\usepackage{comment}

\newcommand{\be}{\begin{equation}}
\newcommand{\ee}{\end{equation}}
\newcommand{\bea}{\begin{eqnarray}}
\newcommand{\eea}{\end{eqnarray}}

\usepackage{amsmath}
\usepackage{amssymb}
\usepackage{mathtools}
\usepackage{chngpage}

\begin{document}
\title{Constraining Axion-Like Particles with HAWC Observations of TeV Blazars}
\author{Sunniva Jacobsen}
\email{sunniva.jacobsen@fysik.su.se, ORCID: orcid.org/0000-0001-8732-577X}
\affiliation{Stockholm University and The Oskar Klein Centre for Cosmoparticle Physics,  Alba Nova, 10691 Stockholm, Sweden}
\author{Tim Linden}
\email{linden@fysik.su.se, ORCID: orcid.org/0000-0001-9888-0971}
\affiliation{Stockholm University and The Oskar Klein Centre for Cosmoparticle Physics,  Alba Nova, 10691 Stockholm, Sweden}
\author{Katherine Freese}
\email{ktfreese@utexas.edu, ORCID: orcid.org/0000-0001-9490-020x}
\affiliation{Department of Physics, University of Texas, Austin, TX 78722, USA}
\affiliation{Nordita, KTH Royal Institute of Technology and Stockholm University, Roslagstullsbacken 23,
10691 Stockholm, Sweden}
\affiliation{Stockholm University and The Oskar Klein Centre for Cosmoparticle Physics,  Alba Nova, 10691 Stockholm, Sweden}

\begin{abstract}
Axion-like particles (ALPs) are a broad class of pseudo-scalar bosons that generically arise from broken symmetries in extensions of the standard model. In many scenarios, ALPs can mix with photons in regions with high magnetic fields. Photons from distant sources can mix with ALPs, which then travel unattenuated through the Universe, before they mix back to photons in the Milky Way galactic magnetic field. Thus, photons can traverse regions where their signals would normally be blocked or attenuated. In this paper, we study TeV $\gamma$-ray observations from distant blazars, utilizing the significant $\gamma$-ray attenuation expected from such signals to look for excess photon fluxes that may be due to ALP-photon mixing. We find no such excesses among a stacked population of seven blazars and 
constrain the ALP-photon coupling constant to fall below $\sim$3$\times$10$^{-11}$~GeV$^{-1}$ for ALP masses below 300~neV. These results are competitive with, or better than, leading terrestrial and astrophysical constraints in this mass range.
\end{abstract}

\maketitle

\section{Introduction}
\label{sec: introduction}

Axions and Axion-like particles (ALPs) are a broad group of spin-0 particles that appear in several extensions of the Standard Model of Particle Physics (SM).  The QCD axion was originally motivated as a particle that arises in the Peccei-Quinn solution to the strong CP problem of QCD ~\cite{PhysRevLett.38.1440, PhysRevLett.40.223, PhysRevLett.40.279, Kim:1986ax}. A broader class of ALPs has been found to generically appear in Kaluza-Klein and superstring theories~\cite{Witten:1984dg, Svrcek:2006yi, Conlon:2006tq, PhysRevLett.97.261802}; for a review see~\cite{Ringwald:2014vqa}. These particles are typically light, yet nonrelativistic, and only weakly coupled to standard model particles, making them suitable dark matter (DM) candidates.

Many ALP searches make use of its coupling to photons, described by the following Lagrangian:
\begin{equation}
\label{eq: ALP-photon mixing lagrangian term}
    \mathcal{L}_{a \gamma}=g_{a \gamma} \mathbf{E}\cdot \mathbf{B} a \ ,
\end{equation}
where $g_{a\gamma}$ is the axion photon coupling constant, $a$ denotes the ALP field strength, $\mathbf{E}$ is the electric field of the photon, and
$\mathbf{B}$ in the strength of an external magnetic field.  Through this coupling, ALPs and photons can convert into each other in the presence of an external $\mathbf{B}$ field.   This mixing can be used to search for axions and ALPs in laboratory experiments~\cite{Sikivie:1983ip, vanBibber:1988ge} and could produce observable effects in astrophysical systems, motivating searches in the Sun, supernovae and gamma-ray bursts~\cite{ Raffelt:1990yz,Berezhiani:1999qh,Csaki:2001yk, Deffayet:2001pc, Raffelt:2006cw,Simet:2007sa}.

Of particular interest for this study, the flux of 
high-energy $\gamma$-rays produced in distant sources and measured on Earth may be significantly modified by photon/axion conversion in the presence of local magnetic fields. 
While \emph{primary} $\gamma$-rays from the sources would be absorbed due to interactions with extragalactic background photons, the ALPs can survive the long journey and then re-convert to \emph{secondary} $\gamma$-rays in the $\mathbf{B}$ field of the Milky Way. 
While such a process is rare because two ALP-photon conversions are necessary, it can produce a detectable excess because very high-energy (VHE,~$\gtrsim$~1~TeV) $\gamma$-rays are severely attenuated by the radiation fields of the intergalactic medium (IGM).  

Thus, if a significant VHE $\gamma$-ray flux is detected, ALPs provide a leading explanation.

Models of photon-photon interactions over intergalactic distances indicate that the extragalactic background light (EBL)~\cite{Hauser:2001xs, Dominguez:2010bv}, which consists of the total radiation from star formation processes and active galactic nuclei (AGN), should be opaque to VHE photons~\cite{1966Natur.211..472J, PhysRev.155.1404}. The attenuation of photons traveling through the EBL is dominated by pair production: $\gamma \gamma_{\rm EBl} \to e^+ e^-$ above a threshold given by:
\begin{equation}
\epsilon_{t h}\left(E_{\gamma}, \theta, z\right)=\frac{2\left(m_{e} c^{2}\right)^{2}}{E_{\gamma}(1-\cos\theta)}
\end{equation}

\noindent where $E_{\gamma}$ is the photon energy and $\theta$ is the angle between the photons. Once this threshold is exceeded, the Universe becomes essentially opaque to VHE photons. This phenomenon suggests that the EBL can be used as a ``screen" to test for axion-photon conversion via an intergalactic ``light through a wall" experiment~\cite{VanBibber:1987rq}.

In this paper, we focus on blazars, which are a class of distant AGN that produce a considerable VHE $\gamma$-ray flux within the accretion disk surrounding the supermassive black hole at their center. AGNs produce two jets perpendicular to the accretion disk, which consist of plasma moving at relativistic speeds. If this jet is oriented towards us, the $\gamma$-ray luminosity is maximized and the AGN is classified as a blazar. Blazars are well-suited for studying the effects of ALPs due to their high $\gamma$-ray flux and strong magnetic fields, which promote axion-photon conversion~\cite{osti_4119469, Blandford:1977ds, Maraschi:1992iz}.

Studies of axion-photon conversion at VHE are, in a sense, simpler than at GeV energies, where searches typically look for an axion-induced ``excess" on top of the primary $\gamma$-ray signal from the blazar itself. Such searches for spectral irregularities have been performed by the Fermi collaboration, which find no evidence for axion-like particles~\cite{Fermi-LAT:2016nkz}. The flux at VHEs can be significantly altered by ALP-photon oscillations in magnetic fields~(\cite{DeAngelis:2007dqd}), a process that can be detected by current and future telescopes such as HAWC and the Cherenkov Telescope Array (CTA).
Studies of the sensitivity of future telescopes to this process have been completed by e.g.~\cite{Mirizzi:2007hr, Sanchez-Conde:2009exi, Meyer&Conrad, Meyer&Conrad&Montanino}. 
In particular, Meyer and Conrad~\cite{Meyer&Conrad} (hereafter, MC14) produced models of the TeV flux from known blazars to predict the portions of parameter space in which upcoming Cherenkov Telescope Array (CTA) observations could constrain the axion parameter space. They included four blazars in their paper and found that for all of these blazars, there were previously untested regions in ALP parameter space where CTA is sensitive to exclude the no ALP hypothesis~\cite{Reesman:2014ova, Meyer:2016wrm}.

In this paper, we focus on observations from the High-Altitude Water Cherenkov Telescope (HAWC). HAWC serves as an optimal instrument for detecting ALP-photon conversion due to its large effective area, sensitivity to a range of photons exceeding $\sim$1-100~TeV, and large field of view~\cite{Abeysekara:2013tza}. Recently, the HAWC collaboration published their third catalogue of TeV $\gamma$-ray sources (3HWC)~\cite{HAWC:2020hrt}. With observations spanning over 4 years and covering $\sim 60\%$ of the sky, reaching a peak sensitivity at declinations of $\sim$20$^\circ$. Importantly, the 3HWC catalog includes an interactive tool, from which the source significance and flux upper limits can be computed for any sky position within the HAWC field of view~\footnote{https://data.hawc-observatory.org/datasets/3hwc-survey/coordinate.php}. Because the 3HWC catalog computes source significances at a standardized energy of 7~TeV, we complement the HAWC data with additional lower-energy data by an ''array'' of Atmospheric Cherenkov Telescopes (ACTs) including VERITAS, H.E.S.S. and MAGIC~\cite{VERITAS:2006lyc, Hinton:2004eu, Lorenz:2004ah}.

In this paper, we calculate the expected $\gamma$-ray flux from 9 blazars in models that either include or do not include an ALP component. The sources have been selected from the Fermi-LAT high energy source catalogue~\cite{Fermi_4FGL}, which provides an unbiased selection of sources due to the completeness of the Fermi-LAT exposure.
We produce a joint-likelihood analysis across all sources, finding no evidence for an ALP component, and set strong upper limits on the ALP-photon mixing parameters, which due to the high-energy sensitivity of HAWC, extend to higher ALP masses than previous $\gamma$-ray studies~\cite{Buehler:2020qsn, Meyer:2020vzy}.

The outline for the paper is as follows. In Section \ref{sec: ALP-photon mixing} we give a brief review of theory behind ALP-photon oscillations in external magnetic fields. In Section~\ref{sec: magnetic field models} we describe the magnetic field models we have used in our calculations of the ALP-photon mixing from the blazar to the Milky Way. In Section \ref{sec: Flux Spectra} we calculate the $\gamma$-ray spectrum from blazars in models with, or without, an ALP component.  In Section \ref{sec: blazar selection} we discuss the properties of the individual blazar sources that have been used in our analysis and what selection criteria have been used. In Section \ref{sec: statistical approach} we present the statistical tests that have been used to determine whether the ALP hypothesis gives a good fit to the observed fluxes from ACTs and HAWC. In Section \ref{sec: results} we present our results and finally in Section \ref{sec: conclusions} we present our conclusions.

\section{ALP-photon Oscillations}
\label{sec: ALP-photon mixing}
In this section, we give a brief review (following ~\cite{Raffelt:1987im,DeAngelis:2011id}) of the theory behind ALP-photon oscillations in the presence of external magnetic fields due to the Lagrangian in Eq.~(\ref{eq: ALP-photon mixing lagrangian term}).

\label{subsec: ALP-photon oscillations}
We study a monochromatic, polarized ALP-photon beam travelling through a cold medium with energy $E$ and wave vector \textbf{k}; for concreteness we take the direction of motion of the beam to be in the $\hat{z}$ direction. The beam has the following equation of motion:
\begin{equation}
\label{eq: eq of motion, ALP-photon beam}
    \left( \frac{d}{d z} + E + \mathcal{M}_0\right)\Psi(z) = 0 \ ,
\end{equation}
where \mbox{$\Psi(y) = (A_1(y), A_2(y), a(y))^T$. Here, $A_1, A_2$} are the photon amplitudes polarized along the $x$- and $y$-axis, $a$ is the ALP field, and $\mathcal{M}_0$ is the mixing matrix. 
The propagation of a generic wave function can be written as: \mbox{$\Psi(z) = \mathcal{T}(z,z_0)\Psi(z_0)$}, where $\mathcal{T}(z,z_0)$ is the transfer matrix that solves Eq.~\eqref{eq: eq of motion, ALP-photon beam} with initial condition $\mathcal{T}(z_0,z_0) = 1$.

Since the photons are propagating through a cold plasma, charge screening produces an effective photon mass which results in a plasma frequency:

\begin{equation}
    \omega_{pl} = \sqrt{\frac{4\pi \alpha n_e}{m_e}} \ .
\end{equation}

Neglecting Faraday rotation, and setting the magnetic field to be pointed along the x-axis, the mixing matrix becomes,
\begin{equation} \mathcal{M}_0 = 
    \begin{pmatrix}
    \Delta_{pl} & 0 & 0 \\
    0 & \Delta_{pl} & \Delta_{a\gamma} \\
    0 & \Delta_{a\gamma} & \Delta_{aa} \\
    \end{pmatrix} \ ,
\end{equation}
where \mbox{$\Delta_{a\gamma} = \frac{1}{2}B g_{a\gamma}$} and \mbox{$\Delta_{pl} = -\frac{\omega_{pl}^2}{2E}$}.
Axion-photon conversion becomes maximal and independent of energy (the strong mixing regime) for $E_{\rm crit} \leq E \leq E_{\rm max}$ where

\begin{equation}
\begin{aligned}
E_{\rm crit} &= \frac{|m_a^2 - \omega_{pl}^2|}{2 g_{a\gamma} B} \ , \\
E_{\rm max} &= \frac{90\pi}{7\alpha} \frac{B_{cr}^2 g_{a\gamma}}{B} \ .
\end{aligned}
\end{equation}
Here \mbox{$B_{\mathrm{cr}} \simeq 4.41 \times 10^{13}$~G} is the critical magnetic field.  At energies above $E_{\rm max}$, ALP-photon oscillations are damped due to QED vacuum polarization.

Photon polarization cannot be measured in the VHE band, thus we focus our analysis on the total photon intensity from summing all polarization states. The beam can be described by a generalized polarization density matrix, $\rho(z) = \Psi(z)\Psi^{\dagger}(z)$, which obeys the following commutator relation:
\begin{equation}
    i \frac{\mathrm{d} \rho}{\mathrm{d}z}=\left[\rho, \mathcal{M}_{0}\right] \ .
\end{equation}
The commutation relation can then be solved as \mbox{$\rho\left(z\right)=\mathcal{T}\left(z, z_0\right) \rho(0) \mathcal{T}^{\dagger}\left(z, z_0\right)$}. 
The photon survival probability (defined as the fraction of photons after mixing in $N$ consecutive domains) for a photon beam, $\rho(0)$, which is initially unpolarized, is given by:
\begin{equation}
    P_{\gamma \gamma}=\operatorname{Tr}\left[\left(\rho_{11}+\rho_{22}\right) \mathcal{T} \rho(0) \mathcal{T}^{\dagger}\right] \ ,
\end{equation}
where $\rho_{11} =\rm diag(1,0,0)$ and $\rho_{22} =\rm diag(0,1,0)$ denotes the polarization along the $x$ and $y$ axes. Given that the initial polarization of
astrophysical blazars in the $\gamma$-ray band is not currently known, we set the initial photon state to be a linear combination of these two polarization states. The magnetic field is not homogeneous in most astrophysical environments. For such fields, the transfer matrix can be split up into N domains, where the magnetic field is treated as constant in each domain. The transfer matrix is then:
\begin{equation}
    \mathcal{T}\left(z_{N}, z_{1} ; \psi_{N}, \ldots, \psi_{1} \right)=\prod_{i=1}^{N} \mathcal{T}\left(z_{i+1}, z_{i} ; \psi_{i} ; E\right) \ ,
\end{equation}
where $\psi_i$ is the angle between the transverse magnetic field in domain $i$ and the polarization state along $z$. Since the mixing matrix also includes photon absorption, $P_{\gamma\gamma}$ also includes absorption in the EBL and reduces to $e^{-\tau}$ when there is no mixing between photons and ALPs ($g_{a\gamma} = 0$).

\section{Magnetic field models}
\label{sec: magnetic field models}
Since the mixing between photons and ALPs only occurs in the presence of an external magnetic field, the magnetic field environment between the photon source and the detector are critical for predicting the expected $\gamma$-ray spectrum at Earth. This includes both magnetic fields from the blazar jet itself, as well as the surrounding intra-Cluster magnetic field for sources which exist near the center of large galaxy clusters. For the blazars we consider in our analysis that are members of a galaxy cluster, the intra-cluster magnetic field is usually dominant. In this section, we give a brief overview of the magnetic field models used in our calculation. We include the blazar jet magnetic field, the intra-cluster magnetic field and the Milky Way magnetic field in our analysis. Note that the two former magnetic fields are the most important for our analysis, since these are blazar-dependent.

\subsection{Blazar Jet Magnetic Fields}
Blazars are AGN whose relativistic jets are pointed directly at the Earth. The simplest jet model was described by Blandford $\&$ Königl~\cite{Blandford:1979za} and has a conical geometry: $R(d) = \phi d$, where $R$ is the radius of the jet, $\phi$ is the opening angle and $d$ is the distance from the base of the jet. Since the jet is pointing towards the Earth, the radiation is blue-shifted. Thus, the photon energy in the comoving frame of of the jet, $E'$, is related to the energy in the lab frame, $E$, by a Doppler factor $\delta_D$:
\begin{equation}
\label{eq: Doppler factor}
\delta_D = \left[\Gamma_L (1 - \beta_j \cos \theta_{\rm obs})\right]^{-1} \ ,
\end{equation}
where $\Gamma_L$ is the relativistic Lorentz factor, $\beta_j = v_j/c$ is the beta factor and $\theta_{\rm obs}$ is the angle between the jet axis and the line-of-sight.\\
The jet magnetic field in the vicinity of the VHE emission zone, as a function of the distance $r$ to the central black hole, can be modelled by its toroidal component,
\begin{equation}
    B^{j} (r) = B_0^j \left(\frac{r}{r_{\rm VHE}}\right)^{-1} \ ,
\end{equation}

where $r_{\rm VHE}$ is the distance from the BH to where the VHE emission occurs. For the blazars we consider in our analysis, $B_0^j$ ranges from \mbox{0.02~--~1}~G and $r_{\rm VHE}$ ranges from \mbox{0.003~--~0.141}~pc. The specific values of these parameters for each source in our analysis is given in Table~\ref{tab:sources overview}.

\subsection{Intra-cluster magnetic field}
In addition to magnetic fields within the blazar jet, blazars that exist within galaxy clusters can include additional fields that facilitate ALP-photon mixing. The magnetic field in galaxy clusters can be modelled as a homogeneous field with Gaussian turbulence -- defined as a field with a (directional) mean of zero and variance $\mathcal{B}^2$. The turbulence follows a power-law $\mathcal{M}(k) \propto k^{q}$ between the minimum and maximum scales of turbulence, $k_L$ and $k_H$. The magnetic field follows the radial dependence of the electron density:
\begin{equation}
    B^{\rm ICM}(r) = B_0^{\rm ICM} \left(\frac{n(r)}{n_0}\right)^{\eta},
\end{equation}
where $r$ is the distance to the cluster center, $n_0$ is the central density and $\eta=0.5$~\cite{Fermi-LAT:2016nkz}. The electron density is parametrized as a $\beta$-profile~\cite{Cavaliere:1976tx, Govoni:2004as}: 

\begin{equation}
n(r) = n_0\left( 1 + (r/r_c)^2\right)^{-\frac{3}{2}\beta_{atm}} \ ,
\end{equation}

\noindent with $r_c$ as the core radius.
The magnetic field strength and electron density are typically of the order of $1~\mu G$ and $10^{-3} \rm cm^{-3}$, respectively. In our calculations, we follow Ref.~\cite{Meyer&Conrad} and set the core radius to $19.33$~kpc for the blazars 1ES0229, PG1553 and PKS1424, which are the sources that lie within large clusters and thus include an additional cluster component to the total $\gamma$-ALP mixing. Because of the random nature of the Gaussian turbulence, it is necessary to simulate a large number of realizations of the magnetic field and investigate the ALP-photon mixing for each of the realizations. In our calculations we have used a total of 300 realizations of the magnetic field, and use the mean value of $P_{\gamma \gamma}$ from these to find the expected flux spectra.

\subsection{Intergalactic Magnetic Fields}
The strength of the intergalactic magnetic field (IGMF) is unknown, but current limits indicate that it is $\geq10^{-16}$ G~\cite{Fermi-LAT:2018jdy}, but $\leq$10$^{-12}$~G~\cite{Sigl:2004yk, Dolag:2004kp}. Notably, the energy density of this magnetic field is much smaller than the Cosmic Microwave Background (CMB) energy density, implying that the IGMF has a negligible effect on particle cooling and attenuation~\cite{Dominguez:2011xy} since losses from ICS of the CMB are more important than synchrotron losses from the magnetic field. 
Moreover, the pair-conversion probability for particle transport over $\sim$Gpc distances still falls far below what one would expect in the magnetic fields in the vicinity of the blazar and in the Milky Way. We do not consider the effect of the IGMF in what follows.

\subsection{Milky Way galactic field}
ALPs can efficiently convert back to photons in the magnetic field of the Milky Way (GMF)~\cite{Simet:2007sa}. In our calculations, we use the same strategy as MC14 and adopt the coherent component of the model described by Jansson and Farrar~\cite{Jansson:2012pc}.
Note that it has been shown in work by e.g. the Planck collaboration~\cite{Planck:2016gdp}, that the random component of the magnetic field is too large in this model. Thus, we are using the modified version of the Jansson~$\&$~Farrar model which takes into account the Planck data, and is included in the GammaALPs package~\cite{Meyer:2021pbp}. We note that unlike searches for diffuse axion fluxes from extragalactic sources~\cite{Vogel:2017fmc}, the fact that we focus on $\gamma$-ray emission from individually known blazars, makes the contribution from the galactic $\gamma$-ray background negligible.

\section{Blazar Spectral Models}
\label{sec: Flux Spectra}
We closely follow the method outlined in Ref.~\cite{Meyer&Conrad&Montanino} to calculate the expected flux spectra from each source at Earth with and without ALP effects. The calculation can be divided into four main steps: First, we pre-fit the blazar spectrum observed by ACTs with a simple-power law:

\begin{equation}
\label{eq: power law}
    \phi_{\rm obs}(E) = N\left(\frac{E}{E_0}\right)^{-\gamma} \ .
\end{equation}

\noindent where we do not include a spectral cutoff because  observations indicate that most blazars do not have an intrinsic spectral cutoff at energies $\sim$1~TeV \cite{Massaro:2005qg, Mazin:2007pn, Dzhatdoev:2021wvv}, though recent observations have indicated that some well-studied blazars such as Mkn 421 and 1ES 1011+496 may show evidence for a spectral cutoff~\cite{deLeon:2019izj, Romoli:2017nrp, Sinha:2016swx}.

Second, the observed spectral data need to be corrected for EBL absorption and possible ALP effects. This is done by dividing the  flux data by the averaged photon survival probability, $\langle P_{\gamma\gamma}\rangle$, in each energy bin. The average photon survival probability, $\langle P_{\gamma\gamma}\rangle$, can be calculated as the energy-integrated average:

\begin{equation}
    \langle P_{\gamma\gamma}\rangle = \frac{\int_{\Delta E} P_{\gamma\gamma}(E)\phi_{\rm obs}(E)}{\int_{\Delta E} \phi_{\rm obs}(E)} \ .
\end{equation}

\noindent We calculate the photon survival probability $P_{\gamma \gamma}$ with and without ALPs by using the GammaALPs package~\cite{Meyer:2021pbp}. 

Third, to find the intrinsic $\gamma$-ray spectrum of each source, we fit a power law of the same type as Eq.\eqref{eq: power law} to the absorption corrected data points. For this portion of the fit, we conservatively only include data points in the optically-thin regime ($\tau < 1$) in order to make our result independent of the exact shape of $P_{\gamma\gamma}$. 
However, in order to make a reasonable fit to the absorption corrected points we must include at least four points, even if a few of these might fall slightly outside of the optically-thin regime. This produces an improved fit to the intrinsic blazar spectrum that takes into account corrections from EBL absorption.

Finally, we compute the expected $\gamma$-ray spectrum on Earth, $\phi_0$, as:

\begin{equation}
\phi_{0}(E) = P_{\gamma\gamma}(E)\phi(E) \ ,
\end{equation}

\noindent where the intrinsic spectrum, $\phi(E)$, is now based on our fit to the optically thin data. The expected spectrum on Earth, $\phi_0$, is similar to the spectrum observed by ACTs in the low-energy regime, but is extrapolated to large energies ($>$1~TeV) where ALP effects are significant. In order to find the total flux in each energy bin as observed by each telescope, we must take into account the energy dispersion between the true, $E$, and reconstructed energy $E'$. The total flux in each bin is then,

\begin{equation}
\label{eq: expected flux, bin}
    \left(\frac{dN}{dE}\right)_i = \int_{\Delta E_i} dE'\mathcal{D}(E, E')\phi_0(E') \ ,
\end{equation}
where $\mathcal{D}(E,E')$ is the energy dispersion which we approximate as a Gaussian with variance $\sigma_E = \rm 0.1E$.

\section{Blazar Selection}
\label{sec: blazar selection}

We select a population of luminous blazars for our analysis as follows.  Each source must fulfill four characteristics:
\begin{enumerate}
    \item The blazar must be included in the \mbox{Fermi-LAT} catalog~\cite{Fermi_4FGL}.
    \item The blazar must be observed in an energy range (typically between \mbox{$\sim$100~GeV -- 1~TeV}), where $\gamma$-ray attenuation is negligible and the primary component is dominant. Such observations have been made by both space-based instruments such as the Fermi-LAT, as well as ground-based Atmospheric Cherenkov Telescopes (ACTs) such as VERITAS, MAGIC and H.E.S.S.. However, we require that the source must have been observed by both Fermi-LAT and a separate ACT. Fermi-LAT typically covers energies in the GeV range, while the ACTs cover energies up to $\sim1$~TeV.
    \item The blazar must be observed in an optically thick regime (in this study $\sim$7~TeV to align with HAWC observations) where the primary $\gamma$-ray flux is suppressed and only the secondary component survives. Following the framework of MC14, we set the minimum optical depth at 7~TeV to be $\tau \geq 4$. Notably, given the higher energy range of HAWC observations (compared to previous studies using VERITAS and MAGIC) - a larger number of candidate blazars exist in this study than in Ref.~\cite{Meyer&Conrad}. 
    \item Finally, we require that the source is not extremely variable -- in order that HAWC observations (taken continuously over the last $\sim$5~years) have observed similar intrinsic fluxes as lower-energy ACT observations (taken sporadically over the last 10~years). This final requirement must be considered carefully -- as the vast majority of blazars are (to some extent) variable. Furthermore, it is possible that ACT observations were scheduled \emph{because} the blazar was flaring in low-energy observations, which would systematically bias our results. 
\end{enumerate}

In addition to these constraints, we require that the source is located at a declination between $-10^\circ$ and $50^\circ$, since the sensitivity of HAWC becomes a factor of two worse outside this region. We also require that the photon flux in the 10~--~500 GeV energy range is larger than $3\times 10^{-11} \rm ~erg s^{-1} cm^{-2}$ and that the $\gamma$-ray spectral index $\Gamma < 2.5$, which indicates that these sources should be detectable in both the GeV and TeV regimes.

Noting that these constraints imply that each blazar is easily observed in the Fermi-LAT energy range, we select sources from the Fermi-LAT high-energy source catalog~\cite{Fermi_4FGL}: we find nine sources that meet all criteria: 1ES 0229+200, PKS 1424+240, PG 1553+113, VER J0521+211, 1ES 1218+304, 3C 66A, 1ES 1011+496, MAGIC J2001+435 and PKS 1222+216. We assume that these sources are not associated with a galaxy cluster, unless this has already been established by MC14 or other references found in the source's TeVCAT page.

Many of the sources have been observed by several ACTs. We base our analysis on the ACT observations that extend to the highest energies. If two observations are similar, we choose the one which has been observed for the longest period of time, since this spectrum will most likely be more similar to what has been observed by HAWC.
Because this represents only a handful of sources -- the peculiarities of each may significantly affect our calculated limit. Thus, we discuss both the physical parameters and multiwavelength observations of each source below.

\subsection{1ES 0229+200}
\label{blzr: 1ES 0229}
The source 1ES 0229+200 was first discovered by H.E.S.S.~\cite{1ES0229_HESS} in 2006. It has been observed by both VERITAS~\cite{1ES0229_VERITAS} and H.E.S.S  with observation times of 46 and 41.8 hours, respectively. The source has a redshift of \mbox{$z = 0.1396$~\cite{Woo_et_al}},
leading us to determine that observations by H.E.S.S extend to an optical depth of \mbox{$\tau \sim 4.6$} while the observations by VERITAS extend to \mbox{$\tau \sim 3.4$}. The HAWC observation at 7~TeV is at an optical depth of 3.15. Since the observations by both H.E.S.S. and VERITAS extend to energies beyond the HAWC observation and have similar exposure times, there is no obvious choice of which spectrum we should use. However, since the H.E.S.S. spectrum has slightly smaller uncertainties and provide an extra data point for our goodness-of-fit analysis, we choose these observations to model the flux spectra at low energies. Note that the spectra observed by VERITAS are comparable with those seen by H.E.S.S. 

The source has also been detected by the Fermi-LAT~\cite{2012ApJ...747L..14V} reaching a significance of 12.5$\sigma$ in the 4FGL catalog~\cite{Fermi_4FGL}. Fermi-LAT observations also indicate that the variability index of this source is only $\sim 6.8$, significantly less than the critical value of 18.5 used by the 4FGL catalog to assign a source as ``variable".
At the time of the H.E.S.S. observations, the source experienced a slight dip in the flux. The flux was $\sim 78\%$ of the mean flux, as observed by the Fermi-LAT. 

According to MC14, the blazar is located in the vicinity of the galaxy cluster WHL 22793 at a redshift offset of only \mbox{$\Delta z = 8\times 10^{-4}$}~\cite{SDSS:2010iou, Wen:2012tm}. The magnetic field strength throughout this cluster is unknown, and therefore we utilize the same strategy as MC14 and employ a modified version of the magnetic field found around the FR I radio galaxy 3C 449, which is also located in a similar galaxy group. The magnetic field parameters used in our simulations are presented in Table~\ref{tab:sources overview}.

    \begin{table*}[!t]

           \begin{center}

        \begin{tabular}{|l|cccccc|} \hline
            \textbf{Source}   & \textbf{1ES 0229} & \textbf{PG 1553}& \textbf{PKS1424}& \textbf{VER J0521}& \textbf{3C 66A} & \textbf{J2001}\\ \hline
             R.A.(J2000) & $02\mathrm{h}32\mathrm{m}53.2\mathrm{s}$&  15h55m44.7s& 14h27m00s &05h21m45s &02h22m41.6s & 20h01m15.6s\\
            Dec.(J2000) & +20d16m21s & +11d11m41s &23d47m40s & +21d12m51.4s& 43d02m35.5s & +43d52m44.4s \\ 
          z & 0.14 & $\geq 0.4$& $\geq 0.6$ & $\geq 0.6$ & 0.34 & 0.18 \\
          Var. index~\cite{Fermi-LAT:2019pir} & 6.799 & 107.2 & 250.7 & 809.0 & 1169 & 1267 \\
            Data & H.E.S.S~\cite{1ES0229_HESS} & VERITAS& VERITAS~\cite{PKS1424_spectrumVERITAS}& VERITAS~\cite{VERJ0521_specVERITAS}& MAGIC~\cite{MAGIC:3C66A}& MAGIC~\cite{MAGIC:J2001}\\\hline
            
           B-field scenario & ICMF+GMF& Jet+ICMF+GMF& Jet+ICMF+GMF & Jet+GMF & Jet+ICMF+GMF & Jet+GMF\\
           $B_0^{Jet} \ \ [\rm G]$ &  & 0.5 & 0.0033 & 0.1 & $0.021$ &  $0.055$\\
            $n_0^{Jet}$ &  & $5.35\times 10^3$ & $10^4$ & $3.15\times 10^3$ & $10^4$ & $5.2\times 10^3$ \\
            $\beta^{Jet}$  &  & 2& 2& 2& 2 & 2\\
            $\delta$ & &35 &30 & 12 & 40 & 27\\
            $\theta_{\rm obs}$ & & $1^\circ$ &$1^\circ$ & $1.8^\circ$& $1^\circ$ & $1^\circ$\\
            $ r_{\rm VHE} \ \ [\rm pc]$ & & 0.063&0.057 & 0.026 & 0.011 & 0.095\\
            $r_{\rm max} \ \ [\rm kpc]$ &  & 1 &1 & 1 &1 & 1\\
            & & & & & & \\
            $B_0^{ICMF} \ \ [\rm \mu G]$ & $3.5$ & 1 & 1 &  &  1 & \\
            $n_0^{ICMF} \ \ [\rm cm^{-3}]$ & $3.7\times 10^{-3}$ & $3.7\times 10^{-3}$& $3.7\times 10^{-3}$&  & $3.7\times 10^{-3}$ &  \\
            $r_{\rm Abell} \ \  [\rm kpc]$ & 100& 100& 100 &  & 100 & \\
            $r_c \ \ [\rm kpc]$ & 19.33 & 19.33& 19.33 &  & 19.33 &  \\
            $\beta^{ICMF}$  &0.42 & 0.42& 0.42&  & 0.42 &  \\
            $\eta$ & 1 & 1 & 1 & & 1 &  \\
            $k_L$  & 0.015& 0.015& 0.015&  & 0.015&  \\
            $k_H$ & 5& 5& 5& & 5& \\
            q & -2.53 & -2.53& -2.53&  &-2.53 &  \\
             \hline\hline
            \textbf{Source} & \textbf{1ES 1218} & \textbf{1ES 1011} & \textbf{PKS 1222} & & & \\ \hline
            R.A.(J2000) &12h21m26.3s& 10h15m04.1s & 12h24m54.4s& & & \\ 
            Dec.(J2000) &+30d11m29s& +49d26m01s &+21d22m46s & & & \\ 
            z & 0.18& 0.21 & 0.43& & & \\
            Var. index~\cite{Fermi-LAT:2019pir} & 60.06 & 267.0 & $2.039\times 10^4$ & & & \\
            Data & VERITAS~\cite{1ES1218_specVERITAS}& MAGIC~\cite{MAGIC:1ES1011+496}& MAGIC~\cite{MAGIC:PKS1222}& & &\\ \hline
            B-field scenario & Jet+GMF& Jet+GMF & Jet+GMF& & & \\
            $B_0 \ \ [\rm \mu G]$ & $2.2\times 10^5$ & $4.8\times 10^4$ &$10^6$ & & & \\
            $n_0 \ \ [\rm cm^{-3}]$ & $10^3$& $7\times 10^2$ & $10^3$& & &\\
             $\beta$  &2 & 2 & 2& & &\\
             $\delta$ & 26 & 26 & 15& & & \\
             $\theta_{\rm obs}$ & $1.4^\circ$ & $1^\circ$ & $1^\circ$ & & & \\
             $r_{\rm VHE} \ \ [\rm pc]$& 0.0031& 0.065 & 0.097& & &  \\
             $r_{\rm max} \ \ [\rm kpc]$ & 1 & 1 & 1& & & \\ \hline
        \end{tabular}
        \end{center}
        \caption{Overview of the sources included in the analysis and the parameters used to model the ALP-photon mixing in their magnetic field. The definition of each parameter and the relevant citations are given in the main text.}
        \label{tab:sources overview}
    \end{table*}
    
\subsection{PKS 1424+240}
\label{blzr: PKS 1424}
The blazar PKS 1424+240 has been observed by both VERITAS~\cite{PKS1424_VERITAS_old, PKS1424_spectrumVERITAS} and MAGIC~\cite{PKS1424_MAGIC}. The redshift of the source is uncertain, but current limits constrain \mbox{$0.6\leq z\leq 1.19$}~\cite{Rovero:2016igo, Yang:2010hh}. Note that some sources set the lower limit of the redshift as low as 0.24~\cite{Prandini:2011zf, Zahoor:2021frv}. In this paper, we set the blazar redshift to 0.6,  which gives us a more conservative limit because it minimizes the value of $\tau$ yielding a smaller difference between the expected fluxes with and without ALP effects. 

The most recent observation by VERITAS~\cite{PKS1424_spectrumVERITAS} (2013) extends to \mbox{$E\sim 500$~GeV}, which corresponds to an optical depth of \mbox{$\tau \sim 4$ for $z = 0.6$}. Using the same redshift, the observation by HAWC at 7~TeV is at an optical depth of 19.8. Observations by MAGIC extend only to \mbox{$E\sim 300$~GeV}. Thus, we will use the flux spectra provided by VERITAS in our analysis.

The source has also been observed by the Fermi-LAT at an extremely high significance of 156$\sigma$. The source has a variability index of 251, which implies that the source does not emit in steady state. However, we stress that the variability index is also strongly proportional to the detection significance -- more luminous sources are more likely to be detected as ``statistically" variable. In this case, the maximum yearly flux (observed in late 2011) only exceeds the mean source flux by a factor of 1.46, justifying our assertion that this source is, at most, moderately variable~\cite{Fermi_4FGL}. The VERITAS observations from 2013 were performed when the source experienced s slight dip in flux. At this time, the flux was about $90\%$ of the mean flux of the source, as observed by the Fermi-LAT. Note that from $\sim$ 2013 to 2019, the source continued to be in a state where the observed flux was below the mean flux.

Studies of the $\gamma$-ray spectrum conducted by Aleksic et al.~\cite{PKS1424_params} found that the emission from PKS1424 was best fit by a two-zone synchrotron self-Compton model with the following best-fit parameters: a magnetic field strength \mbox{$B = 0.033$~G}, Doppler factor \mbox{$\delta = 30$} and a radius of \mbox{$4.8\times 10^{16}$}~cm of the VHE plasma emitting blob. Assuming an angle between the jet and the line of sight of \mbox{$\theta = 1^o$}, this gives a bulk Lorentz factor of \mbox{$\Gamma \sim 16$}. An analysis by Ref.~\cite{Rovero:2016igo} has found that there is a $98\%$ chance that the blazar is hosted by a group of galaxies located at a redshift of \mbox{$z = 0.6010$}. Since the magnetic field in this environment is unknown, we assume that it can be described by the same parameters as the intracluster magnetic field surrounding 1ES 0229. We will include ALP/photon mixing in both the jet of the AGN and in the intracluster magnetic field in our calculations.

\subsection{PG 1553+113}
\label{blzr: PG1553}
The blazar PG 1553 has been observed by VERITAS~\cite{Aliu:2014kca}, MAGIC~\cite{MAGIC:2011sam} and H.E.S.S~\cite{HESS:2006cui, HESS:2007gtr}. It has also been detected by the Fermi-LAT~\cite{Fermi-LAT:2009hku}, with a significance of 175$\sigma$ in the 4FGL catalog.
It was observed by H.E.S.S. for 7.6 hours in 2005 and 17.2 hours in 2006. The source has also been observed by VERITAS for 95 hours and by MAGIC for 28.7 hours. The observations by VERITAS and MAGIC extend to energies up to ~0.5 TeV, corresponding to an optical depth of $2.6$ assuming a redshift of 0.43. With the same redshift, the HAWC observation at 7~TeV is at an optical depth of $12.6$. The 2006 observations by H.E.S.S. extend to energies up to ~1 TeV, but are subject to large uncertainties on the flux above 0.45~TeV. Since the H.E.S.S. datapoints with smaller uncertainties are at similar energies to the VERITAS and MAGIC data, we choose to use the data from VERITAS since this is the most recent and has the longest exposure. Since HAWC operates continuously over a 5-year period, the spectrum should be more similar to what has been observed at HAWC with a longer exposure time.
At the time of the VERITAS observations, the source experienced slight dip in the flux with $\sim 88\%$ of the mean flux, as observed by the Fermi-LAT.

According to the Fermi variability analysis~\cite{Fermi-LAT_variability}, it has a variability index of $\sim107$, which (similar to PKS 1424+240) corresponds to a source that is somewhat variable. In the case of PG 1553+113, the variation between the maximum yearly flux (obtained in late 2017/early 2018) and the mean $\gamma$-ray flux is only 1.23) indicating that flux variability likely has a relatively modest effect on our results.

The distance to this object is uncertain. Danforth et al.~\cite{Danforth:2010vy} found in 2010 that a lower limit can be set at $z \leq 0.4$, and recent determinations performed by Refs.~\cite{HESS:2015ygw, Nicastro:2018iya, Jones:2021aph} seem to prefer a redshift in the range $0.4-0.5$. Thus, we adopt the best-fit value from Refs.~\cite{ Nicastro:2018iya, Jones:2021aph} and set the redshift to be $z = 0.43$.

Ref.~\cite{MAGIC:2011sam} models the SED using a one-zone synchrotron model, and find the following best-fits for the parameters that describe the VHE emitting region: \mbox{$B = 0.5$~G, $R = 10^{16}$~cm and $\delta = 35$}.
By performing a cross-correlation of the galaxy cluster catalogs GMBCG and WHL~\cite{SDSS:2010iou, Wen:2012tm}, MC14 found that PG 1553 is in the vicinity of a galaxy cluster consisting of 8 member galaxies. Following MC14, we use the same magnetic field parameters for this environment as 1ES0229 and conservatively fix the field strength to \mbox{$\rm B = 1~\mu G$}.

\subsection{VER J0521+211}
\label{blzr: VERJ0521}
The intermediate-peaked BL Lac object VER J0521 was first discovered in 2009 and has now has been observed by VERITAS~\cite{VERJ0521_data} for 14.5 hours. The redshift of this source is still unknown, but a lower limit can be set at \mbox{$z > 0.18$}~\cite{VERJ0521_data, Paiano:2017pol}. The VERITAS observations extend to $\sim 1.1$~TeV, corresponding to an optical depth of $1.7$ using the lower limit of the redshift. The HAWC observation at 7~TeV is at an optical depth of $4.2$. Observations in the 4FGL catalog detect this source at a significance of 120$\sigma$ and with a variability index of 809, which makes this source somewhat more variable than PKS 1424+240 and PG 1553+113. The maximum flux was observed in mid-2013 and had a flux that was 1.9 times higher than the average source flux within the 4FGL catalog. 
At the time of the VERITAS observatios, the source experienced a slight dip in the flux with a flux of $\sim 73\%$ of the mean flux, as observed by the Fermi-LAT. Note that there was a similar dip in the flux at the time of the HAWC observations of the source. 

A two-zone SSC model of the SED was performed by the MAGIC collaboration in 2020~\cite{VERJ0521_params}. The zone responsible for the VHE emission has the following best-fit parameters: \mbox{$\rm B = 0.1~G $, $\delta = 12$ and radius $\rm R = 1.3\times 10^{16}$~cm}. Given a viewing angle of \mbox{$\theta = 1.8^o$} and a simple, conical jet geometry,  the distance between the VHE emitting region and the jet is \mbox{$\rm r_{VHE} = 0.026$~pc}. 

The source has not been explicitly associated with a galaxy cluster and thus we will omit any contribution to the ALP-photon mixing from an intra-cluster magnetic field.

\subsection{1ES 1218+304}
\label{blzr: 1ES1218}
The BL Lac object 1ES 1218 has been observed by VERITAS~\cite{1ES1218_VERITAS_spectrum} and MAGIC~\cite{MAGIC:2006xcg} for 27.2 and 8.2 hrs, respectively.  Note that observations by MAGIC were also performed in 2010 and 2011 during an observation campaign for 1ES 1215+303, where the source always stayed in the field-of-view. However, no energy spectrum is publicly available from this observation campaign~\cite{1ES1218_MAGIC}.
Observations by VERITAS extend to \mbox{$\sim 1.8$}~TeV while the observations from MAGIC extend to \mbox{$\sim 0.6$}~TeV. This corresponds to an optical depth of \mbox{$\tau \sim 2.1$ and $\tau \sim 1.1$} respectively, with a redshift of \mbox{$z = 0.182$}. The HAWC observation at 7~TeV is at an optical depth of $4.3$. Since the observations from VERITAS extend to higher energies and are taken over a longer period of time, we will use these in our calculations. Note that a flare was detected during the observations performed by VERITAS that lasted over two days, making up $~11\%$ of the total live-time. However, changes to the VHE photon index during the flare are statistically insignificant.
At the time of the VERITAS observations, the source experienced an increased flux, about $132\%$ of the mean flux as observed by the Fermi-LAT. Note that at the time of the HAWC observations, a similar increase in the flux was observed measuring up to $144\%$ of the mean flux. 

The source has been observed in the GeV range by the Fermi-LAT, with 4FGL observations reaching a significance of 69$\sigma$ and a variability index of 60. The maximum annual flux from the source (reached in mid-2018) is roughly 1.4 times brighter than the average source flux.

Singh et al.~\cite{1ES1218_jetparams} found that the zone responsible for emission of VHE photons can be described by a one-zone SSC model with the following parameters: \mbox{$\rm B = 0.22$~G, $\rm R = 2.2\times 10^{15}$~cm}, \mbox{$\delta = 26$} and a viewing angle of $1.4^o$. 

The source has not been explicitly associated with a galaxy cluster and thus we will conservatively omit any contribution to the ALP-photon mixing from an intra-cluster magnetic field.

\subsection{3C 66A}
\label{blzr: 3C66A}
The intermediate-frequency peaked BL Lac object 3C 66A has been observed by several telescopes, e.g. VERITAS~\cite{3C66A_specVERITAS} and MAGIC~\cite{MAGIC:3C66A}. The blazar is located within a galaxy cluster at \mbox{$z = 0.34$}~\cite{3C66a:GalaxyCluster}. Using this redshift, observations by MAGIC extend up to an optical depth of \mbox{$\sim 1.5$ ($E \sim 0.4$~TeV)} and the observations by VERITAS up to \mbox{$\sim 1.72$ ($E \sim 0.45$~TeV)}. The HAWC observation at 7~TeV is at an optical depth of \mbox{$~9.3$}. The observations by VERITAS were performed when the blazar was in a flaring state. Therefore, we will use the observations by MAGIC to model the low-energy data.
At the time of the MAGIC observations, the source experienced an increased flux about $1.2$ times larger than the mean flux as observed by the Fermi-LAT. Note that at the time of the HAWC observations, the source experienced a dip in the flux with a minimal flux $50\%$ of the mean flux.

Fermi-LAT observations in the GeV range detect this source at a significance of 170$\sigma$ with a high variability index of 1170. The maximum annual flux of this source (reached in early 2009) was 1.7$\times$ brighter than the average source flux.

The SED of the source was modelled by Ref.~\cite{3C66a:SEDmodel}. They model the SED using both a pure SSC model and a SSC+EC model. We will use the best-fit parameters from the SSC+EC model since these models allow for variability timescales in better agreement with what has been reported. The parameters describing the emission region for this model with \mbox{$z = 0.3$} are: \mbox{$\rm R = 1.5\times 10^{16}$~cm, $\delta = 40$ and $\rm B = 0.21$~G}. We assume the viewing angle is $1^o$. The magnetic field of the galaxy cluster is unknown, so again we use the same strategy as MC14 and use the modified version of the magnetic field found around 3C 449. 

\subsection{1ES 1011+496}
\label{blzr: 1ES1011}
The high-frequency peaked BL Lac object 1ES 1011+496 has been observed by both VERITAS~\cite{VERITAS:1ES1011+496} and MAGIC~\cite{MAGIC:1ES1011+496}. It has been observed by VERITAS for 10.4 hours and by MAGIC for 19.4 hours (after quality cuts). We will use the observations from MAGIC since they extend over the largest period of time.  The flux observed by MAGIC extends up to \mbox{$\sim0.75$}~TeV, or an optical depth of \mbox{$\tau \sim 1.6$} using a redshift of \mbox{$z = 0.212$}. The HAWC observation at 7~TeV is at an optical depth of ~$5.2$.

Fermi-LAT observations detect this source at a significance of 158$\sigma$ in the 4FGL catalog, and with a variability index of 260. The maximum annual GeV $\gamma$-ray flux (observed in late 2014) reaches an amplitude 1.4 times brighter than the average $\gamma$-ray flux over the full observation period.
At the time of the MAGIC observations, the source experienced a slight dip in flux about $82\%$ of the mean flux, as observed by the Fermi-LAT.

The MAGIC collaboration has also performed an analysis of the multi-wavelength spectrum of 1ES 1011+496 and used this to model the SED of the source~\cite{MAGIC:1ES1011+496}. They have found that the SED can be described by a one-zone SSC model with the following parameters: $\rm B = 0.048$~G, $\delta = 26$, $R = 3.25\times 10^{16}$~cm. We assume that the viewing angle is $1^{o}$, which will be our default viewing angle.

The source has not been explicitly associated with a galaxy cluster and thus we will conservatively omit any contribution to the ALP-photon mixing from an intra-cluster magnetic field.

\subsection{MAGIC J2001+435}
\label{blzr: J2001}
The high-frequency peaked Bl Lac object MAGIC J2001+435 was first discovered by the Fermi-LAT in 2009~\cite{FERMIJ2001} and later observed by MAGIC in 2009 and 2010~\cite{MAGIC:J2001}. Observations by MAGIC extend up to energies of $0.4$~TeV or an optical depth of $\sim 0.65$, using a redshift of $z = 0.18$. The HAWC observation at $7$~TeV is at an optical depth of $4.2$.

Fermi-LAT observations detect this source at a significance of just 58$\sigma$, but with a variability of 1270, indicating that this is a highly variable source. The maximum flux, observed during MAGIC observations in early 2010, was a remarkable 3.5$\times$ brighter than the average source flux. We note that the variation is more extreme, as this bright flare contributes 1/12 of the average flux. Based on this assessment, we remove this source from our analysis and do not analyze it further.

\begin{figure*}
\begin{adjustwidth}{-\oddsidemargin-0.85in}{-\rightmargin}
    \includegraphics[width=0.9\paperwidth]{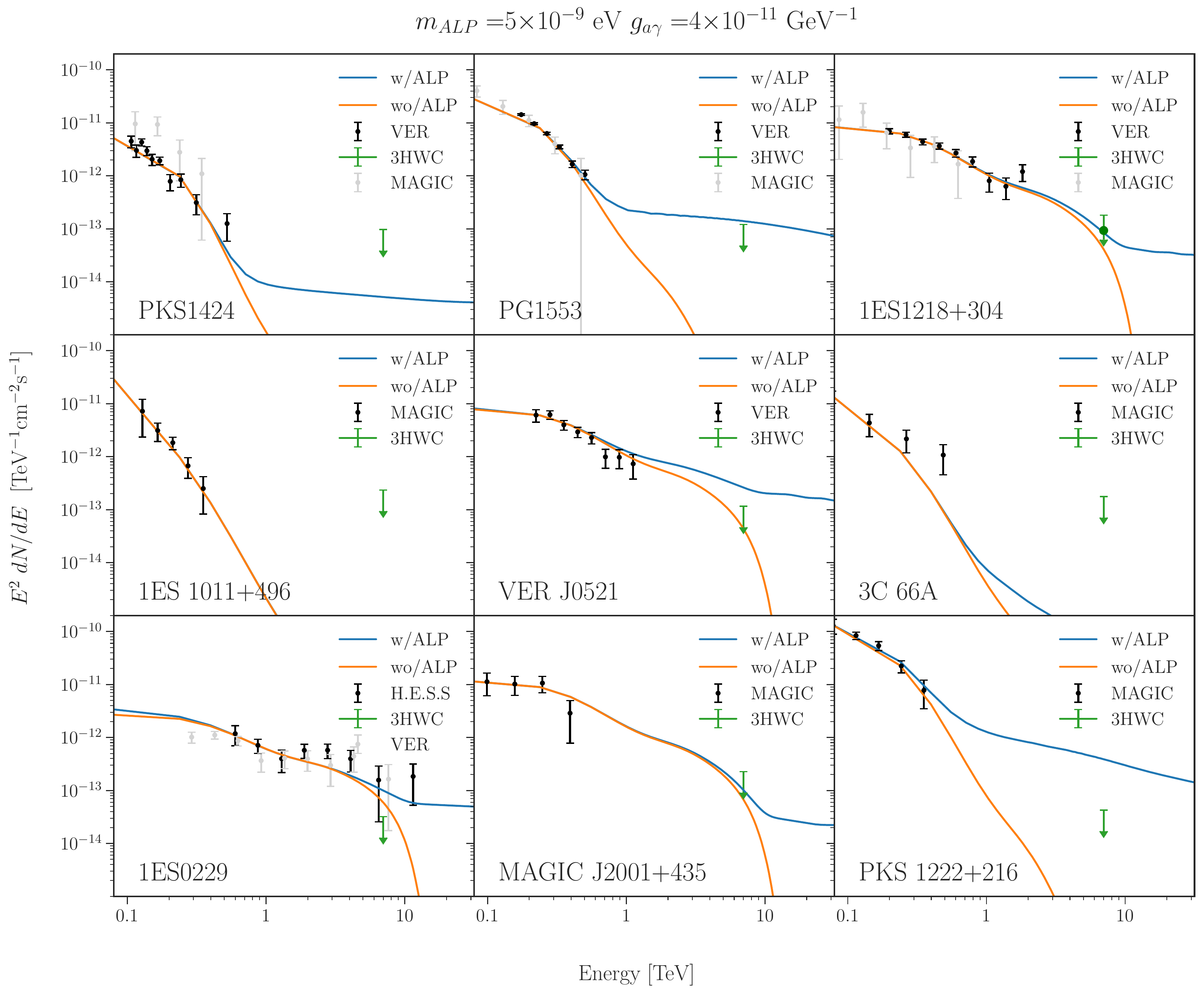}
    
    \caption{ Simulated flux spectra as observed at Earth with an ALP (blue line) and without (orange line). The flux with a contribution from ALPs has been simulated for an ALP with mass $m_a = 5\times 10^{-9}$~eV and coupling strength to photons of $g_{a\gamma} = 4\times 10^{-11}~\rm GeV^{-1}$. The black points show the observed flux spectra as observed by various telescopes, while the green data point show the HAWC-3 upper limit at the source location. For the sources where we had two different sets of comparable observations, we also show the data (denoted by the gray data points) that we \emph{did not} use in our calculations. We note that the sources MAGIC J2001+435 and PKS 1222+216 were observed by ACTs during blazar flares, and while shown here, are not included in any subsequent analysis. Some sources (e.g., 1ES 1011+496) are best fit with relatively soft spectral indices in the sub-TeV band, making them unsuitable candidates for ALP studies at 7~TeV.}
    \label{fig:flux spectra}
\end{adjustwidth}
\end{figure*}

\subsection{PKS 1222+216}
\label{blzr: PKS1222}
PKS 1222+216 represents the only Flat-Spectrum Radio Quasar (FSRQ) we will use in our analysis.
Very high-energy emission from this source was first discovered by MAGIC in 2010 during a short (0.5 hours) observation campaign~\cite{MAGIC:PKS1222}. These observations coincide with lower-energy activity measured by Fermi/LAT~\cite{Fermi:PKS1222}. The observations performed by MAGIC extend up to energies of $\sim 0.36$~TeV, or an optical depth of $\tau \sim 1.8$, using a redshift of $z \sim 0.43$. Note that these observations were made when the source was in a flaring mode. The HAWC observation at 7~TeV is at an optical depth of $12.6$.\\

The emission region of the source has not been modelled completely, but it is constrained due to the fast variability of the signal. The MAGIC collaboration has found that the source has fast variability $t_{var} \sim 10$min, indicating an extremely compact emission region. They find a lower limit on the Doppler factor of $\delta > 15$ and the distance of the emitting region to be $d > \rm R_{BLR} = 3\times 10^{17}$~cm. Assuming a simple conical jet geometry, this gives $R \sim 4\times 10^{16}$~cm. Furthermore, we take a conservative estimate of the magnetic field strength of $\rm B = 1$~G. 

The source has not been explicitly associated with a galaxy cluster and thus we will omit any contribution to the ALP-photon mixing from an intra-cluster magnetic field.
Since this source has only been observed in a flaring mode, we cannot compare the expected fluxes directly to the HAWC data. Thus, we will not include this source in the joint likelihood analysis.

\begin{figure*}
\begin{adjustwidth}{-\oddsidemargin-0.7in}{-\rightmargin}
    \includegraphics[width=0.9\paperwidth]{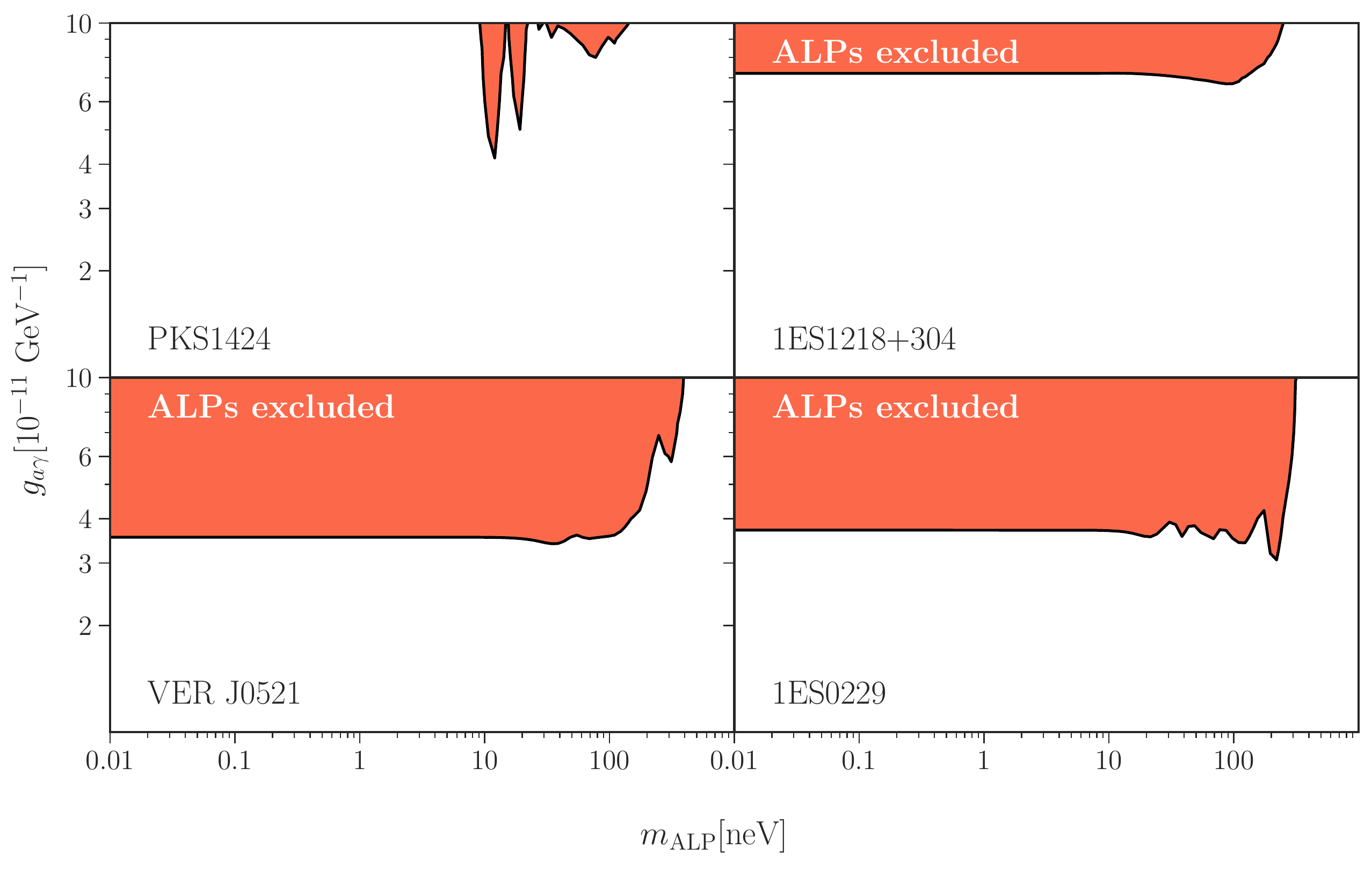}
    \caption{The regions of ALP parameter space where the ALP hypothesis is disfavoured at $>3\sigma$ compared to the no-ALP hypothesis for four individual sources. For the sources 1ES 1011+496, PG1553+113 and 3C 66A, there are no regions of the parameter space shown here where one can distinguish between the ALP and no ALP hypothesis at a level of $>3\sigma$, while the sources MAGIC J2001+435 and PKS 1222+216 are not included in this part of the analysis because the ACT observations were taken during blazar flares.}
    \label{fig:constraints on ALP parameter space}
\end{adjustwidth}
\end{figure*}

\section{Statistical Approach}
\label{sec: statistical approach}
ALP-photon mixing increases the transparency of the universe for very-high energy $\gamma$-rays, increasing the expected TeV flux from blazars. This implies that the signal of ALP-photon mixing is given by a flattening of the $\gamma$-ray spectrum -- as the VHE $\gamma$-ray flux shifts from an exponentially falling spectrum (produced by $\gamma$-ray attenuation) to a flat-spectrum given by a relatively energy-independent ALP-photon mixing parameter that matches the injected $\gamma$-ray spectrum from each blazar.

Using the combined ACT and HAWC measurements of each source, we compare the goodness-of-fit of the ALP vs no ALP hypothesis by using the test statistic: 
\begin{equation}
    TS = 2\left(\mathcal{L}_{\rm ALP} - \mathcal{L_{\rm no ALP}}\right).
\end{equation}
In the absence of ALP effects, the test statistic will be distributed according to a $\chi^2$ distribution with two degrees of freedom since models with an ALP add two more free parameters, $m_{\rm ALP}$ and $g_{a\gamma}$.  Employing Wilks' theorem, we can interpret the log-likelihood ratio by comparing the $\chi^2$ fits of our ALP and no-ALP models to the combined ACT and HAWC datasets as:

\begin{equation}
    \mathcal{L}_{\rm obs} = \sum_{i=1} \frac{(f_{i, \rm exp}(m_{\rm ALP}, g_{a\gamma}) - f_{i, \rm obs})^2}{\sigma_i^2} \ ,
\end{equation}
where $f_{i,exp}$ is the expected flux in energy bin $i$, as found by Eq.\ref{eq: expected flux, bin} and $f_{i, obs}$ is the flux observed by ACTs. $\sigma_i$ denote the errors in the observed flux spectra. \\
HAWC quotes the $2\sigma$ upper limit of their flux measurements, along with the best-fit fluxes. Thus, the contribution to the total log-likelihood for the expected flux at 7~TeV is given by a modified $\chi^2$:
\begin{equation}
\begin{aligned}
    \mathcal{L}_{\rm HAWC} =  \frac{2(f - f_{BF})^2}{(f_{ul} - f_{BF})^2},
\end{aligned}
\end{equation}
where $f_{BF}$ and $f_{ul}$ are the best-fit flux and the upper limit on the flux as quoted by HAWC.

\begin{figure}
    \centering
    \includegraphics[width=0.41\paperwidth]{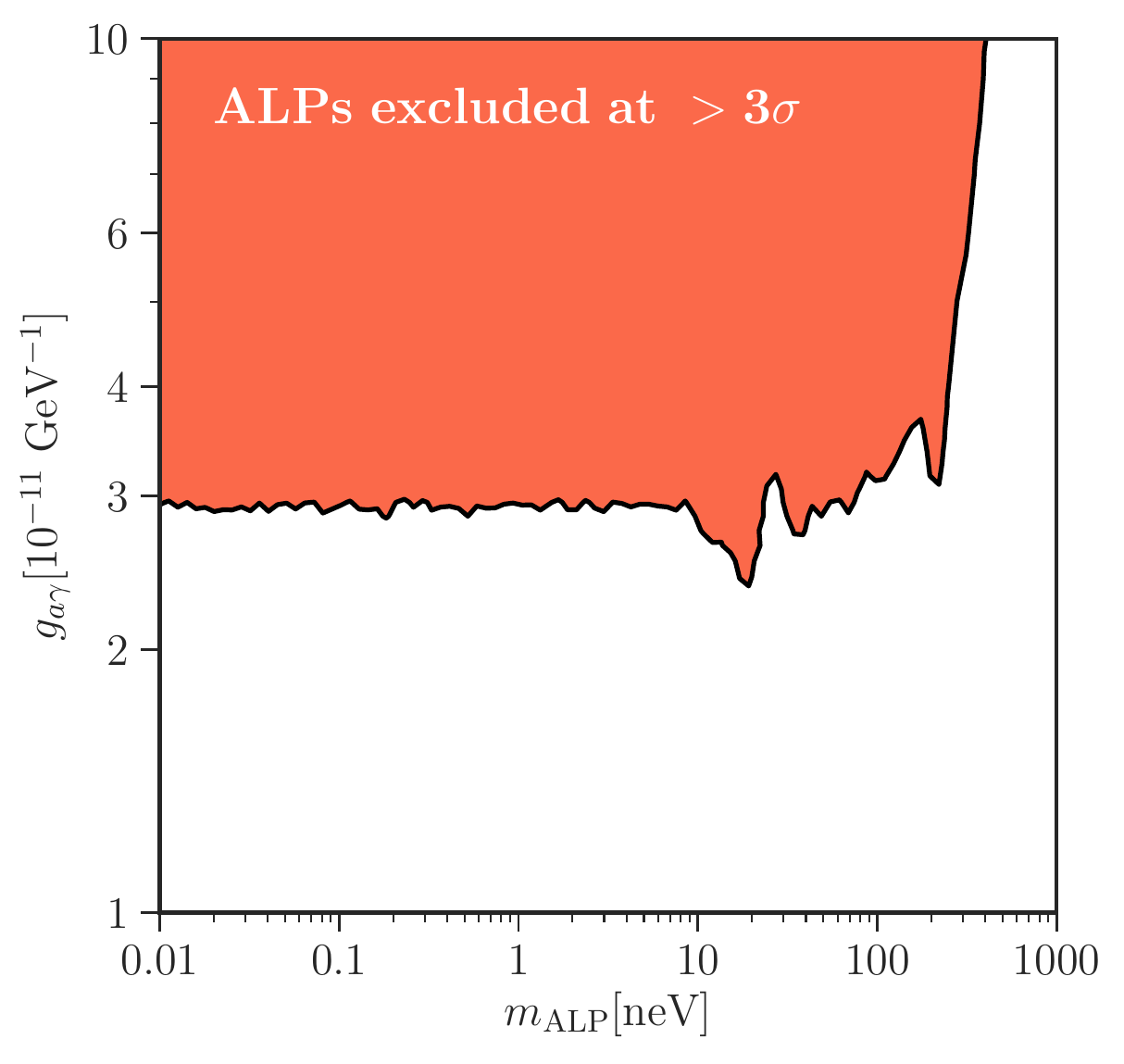}
    \caption{The filled contour shows the parts of the ALP parameter space where the ALP hypothesis is disfavoured at $>3\sigma$ after performing a joint likelihood analysis. Note that the sources J2001 and PKS1222 are not included in this analysis since they have only been observed in flaring states, making them unsuitable for comparison with the HAWC data.}
    \label{fig: stacked constraints}
\end{figure}

\section{Results}
\label{sec: results}

In Figure~\ref{fig:flux spectra}, we compare the expected flux (as calculated in Sec.~\ref{sec: Flux Spectra}) from each blazar with the combined ACT and HAWC datasets. The effect of ALP-photon mixing on the expected $\gamma$-ray flux from each blazar varies due to the different magnetic field environments in the vicinity of the blazar, the redshift of each blazar source, and the assumed ALP parameters.  We show results for models that both include, and do not include, an ALP. For the ALP models, we adopt an assumed mass of $m_{\rm ALP} = 5$~neV and ALP-photon mixing of $g_{a\gamma} = 4\times 10^{-11},$ values chosen to be relatively close to the 3$\sigma$ upper limits on the ALP parameter space that will be obtained by our study.

Figure \ref{fig:flux spectra} immediately demonstrates several key results: (1) the effect of ALPs on the $\gamma$-ray spectra typically become apparent at energies above 1~TeV, with the exact transition depending on the redshift of each source. This demonstrates that, between the energy ranges of ACT and HAWC observations, our models efficiently transition from the regime where the ``primary" $\gamma$-ray flux (produced by unattenuated $\gamma$-rays from each blazar) is dominant to the ``secondary" regime where the $\gamma$-ray flux from ALP-photon conversion dominates, (2) the spectrum in the ``secondary" regime is much harder than the spectrum in the primary regime due to the fact that $\gamma$-ray attenuation is no longer affecting the signal, indicating both that spectral signatures can provide important information regarding axion properties, and also that our results are relatively unaffected by energy-dispersion or energy-calibration uncertainties in the HAWC energy range (3) HAWC observations for all sources present only upper limits, and no source has been detected in the HAWC energy band. The significances of the sources as detected by HAWC are in the range $\sqrt{TS}$~=~-1.6 -- 2.24, where 1ES 1218+304 is the most significant source. We note that a negative value for $\sqrt{TS}$ is interpreted as a positive test statistic for a source with a negative flux.

We can already observe that this combination of $(m_{ALP}, g_{a\gamma})$ is constrained by the sources 1ES 0229, VER J0521 and PKS 1222. In these cases, the HAWC upper limit on the flux falls below the expected blazar flux in models with ALPs. For sources such as PKS1424, PG1553, 1ES 1218+304, 3C 66A and MAGIC J2001+435, the expected fluxes in models with and without ALPs both fall below the HAWC sensitivity, preventing us from using these sources to exclude the ALP hypothesis for this combination of $(m_{ALP}, g_{a\gamma})$. We note that some of these sources may be capable of constraining the ALP parameter space with upcoming instruments such as CTA, as discussed in MC14.

In Figure~\ref{fig:constraints on ALP parameter space}, we show the regions of the $(m_{\rm ALP}, g_{a\gamma})$ parameter space for our four most constraining sources, where the ALP hypothesis is disfavoured by more than $3\sigma$ compared to the no ALP hypothesis. In these regions, the upper limit from HAWC is consistent with the expected flux without ALP effects, but inconsistent with the flux including ALP effects. We do not include a likelihood analysis of the sources MAGIC J2001+435 and PKS 1222+216 because the ACT observations of each source were recorded during blazar flares, which makes them unsuitable for comparisons with the HAWC upper limits, which were based on continuous observations spanning several years. Because our modeled HAWC fluxes are normalized to ACT observations, our model will produce spuriously strong limits in the HAWC band. We note, however, that future ACT observations during non-flaring states may produce strong limits from these blazars.

\begin{flushleft}
\begin{figure}[t!]
    
    \includegraphics[width=0.40\paperwidth]{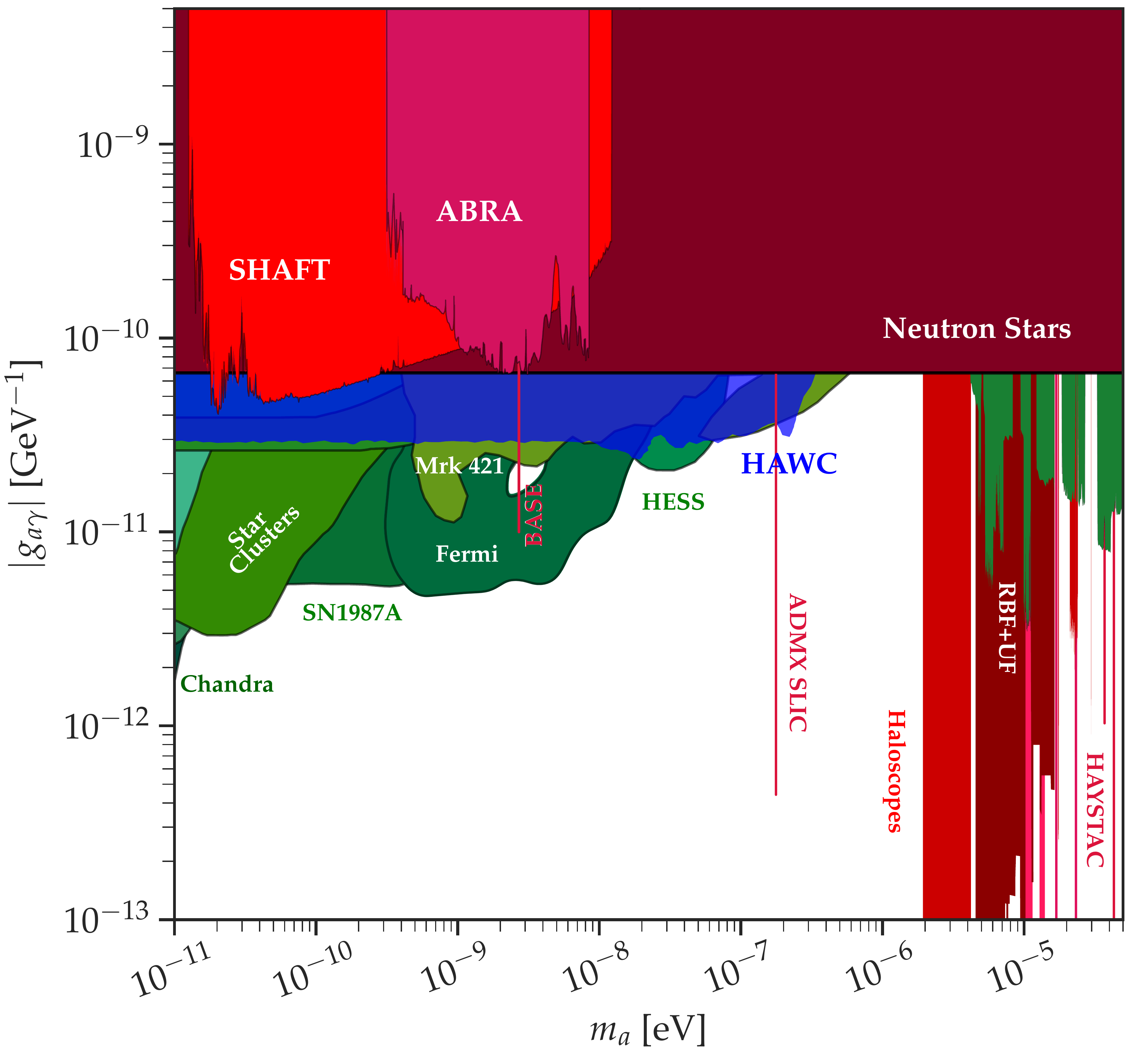}
        \caption{Comparison of existing limits on the ALP parameter space and the limits derived in this paper. This plot, along with the existing limits, has been made using the AxionLimits package from Ref.~\cite{ciaran_o_hare_2020_3932430}. The limits derived in this paper are shown in blue and denoted by "HAWC". The existing limits from other experiments are derived from Refs.~\cite{Helio1, helio2, helio3, helio4, helio5, helio6, helio7, helio8, chandra1, chandra2, chandra3, chandra4, Nstars1, Nstars2, Nstars3, SN19871, SN19872, SN19873, StarClusters, CAST:2017uph, Gramolin:2020ict, Salemi:2021gck, HESS:2013udx, Fermi-LAT:2016nkz, Li:2020pcn} }
    \label{fig: existing limits and ours}
\end{figure}
\end{flushleft}

We note that our theoretical models show that there are significant differences between the $\gamma$-ray fluxes expected in models with and without ALP effects, even for sources and regions of parameter space where the ALP-induced flux falls below HAWC sensitivity. 
This is illustrated in Figure~\ref{fig:flux spectra} by the sources PG1553+113, PKS1424+240, 3C 66A, MAGIC J2001+435 and 1ES1218+304. In these cases, upcoming observations (e.g., by HAWC, LHAASO, or the Cherenkov Telescope Array), could be used to distinguish between the fluxes with and without ALP effects.

Figure~\ref{fig: stacked constraints} shows the combined constraints on the ALP parameter space, where the filled contour denotes the regions of parameter space excluded at $>3\sigma$. We find that the ALP hypothesis is excluded at $>3\sigma$ for $\rm g_{a\gamma} \gtrsim 3\times 10^{-11} ~{\rm GeV}^{-1}$ and masses up to $\sim 200-300 \times 10^{-9}$ eV.  Figure~\ref{fig: existing limits and ours} shows how these limits compare to previous results. As can be seen, the limits derived in this paper are comparable to previous limits in the ALP parameter space.  We have managed to significantly constrain smaller couplings than {\it e.g.} CAST~\cite{CAST:2017uph}, SHAFT~\cite{Gramolin:2020ict} and ABRA~\cite{Salemi:2021gck}, and a larger ALP mass than other astrophysical searches such as H.E.S.S.~\cite{HESS:2013udx} and Fermi~\cite{Fermi-LAT:2016nkz}. Note that the higher-mass portion of our limits are similar to those recently derived in Ref.~\cite{Li:2020pcn}, which are based on searches of ALP-photon oscillations in the spectrum of Mrk 421 with ARGO-YBJ, but are derived from a larger sample of astrophysical sources.

\section{Conclusions}
\label{sec: conclusions}
In this paper, we have studied how an ALP component alters the TeV $\gamma$-ray flux expected from a number of known blazars. Due to pair-production interactions with the EBL, the $\gamma$-ray flux from distant sources is expected to be strongly attenuated at TeV energies. This effect can be decreased if $\gamma$-rays can resonantly convert into ALPs that can survive the passage through the EBL. Specifically, since ALPs can mix with photons in the presence of an external magnetic field, photons from an extragalactic source may mix into ALPs in the magnetic fields surrounding the source, travel unattenuated through the EBL and then mix back into photons in the Milky Way magnetic field. It has been shown in previous work that upcoming telescopes such as CTA may be sensitive to the differences in the expected spectra of blazars with and without ALP effects~\cite{Meyer&Conrad}.

We have studied whether this effect can be used to constrain the parameter space of ALPs using a combination of data from \emph{existing} ACTs and the HAWC telescope. We compare the expected flux centered at the 7~TeV energy-range corresponding to HAWC observations in models that do or do not include an ALP component. We find that the expected flux with an ALP component is excluded at $>3\sigma$ for ALP-photon couplings larger than $~3\times 10^{-11}~\rm GeV^{-1}$ and masses smaller than $~300\times 10^{-9}$~eV. While these limits fall short of constraints from Fermi-LAT observations for axion masses $\sim$10$^{-9}$~eV, the higher energy range of HAWC observations make these studies comparable with the best limits derived from laboratory and astrophysical searches in the mass range near 10$^{-7}$~eV, as shown in Fig.~\ref{fig: existing limits and ours}. 

We note that these existing limits fall just short of the anticipated limits for upcoming CTA observations, as calculated by MC14 ($~2\times 10^{-11}~\rm GeV^{-1}$ for axion masses smaller than $~100\times 10^{-9}$~eV). Upcoming HAWC observations (including both two recent years of data not included in the 3HWC catalog and improved spectral reconstructions) may strengthen these limits by a factor of a few over the next few years.

Additionally, and similarly to MC14, we have found that ALPs may produce significant differences in the high-energy $\gamma$-ray spectrum of multiple blazars, which provides significant theoretical potential for our $\gamma$-ray searches, even if current instrumentation do not yet reach the sensitivity to probe these spectral differences. Thus, future telescopes with higher sensitivity than HAWC, such as the CTA and LHAASO may be able to constrain larger parts of ALP parameter space using the method described in this paper. 

Finally, we note that improved observations and models of $\gamma$-ray emission from blazars, including better measurements of their time-averaged $\gamma$-ray flux and spectrum (including the potential for high-energy cutoffs in the intrinsic blazar spectrum), as well as improved modeling of their magnetic fields, viewing angles, and surrounding cluster magnetic fields, may significantly improve our axion limits. Moreover, the addition of similar instrumentation in the southern hemisphere (such as SWGO~\cite{BarresdeAlmeida:2020hkh}) would significantly increase the available blazar sample. \newline

\vspace{1cm}

\begin{acknowledgments}
    We would like to thank Manuel Meyer for making the GammaALPs package, which was employed in this analysis, publicly available. We would also like to thank Ciaran O'Hare for making the AxionLimits package, which was employed to produce Fig. \ref{fig: existing limits and ours}, publicly available. We would also like to thank Pierluca Carenza, J. Patrick Harding and Alexander J. Millar for helpful comments regarding the manuscript. S.J. and K.F. acknowledge support by the Vetenskapsr{\aa}det (Swedish Research Council) through contract No.  638-2013-8993 and the Oskar Klein Centre for Cosmoparticle Physics.  TL is partially supported by the Swedish Research Council under contract 2019-05135, the Swedish National Space Agency under contract 117/19 and the European Research Council under grant 742104. K.F. is Jeff \& Gail Kodosky Endowed Chair in Physics at the University of Texas at Austin, and is grateful for support. K.F. acknowledges funding from the U.S. Department of Energy, Office of Science, Office of High Energy Physics program under Award Number DE-SC0022021 at the University of Texas, Austin.

\end{acknowledgments}

\bibliography{ALP_hawc.bib}
\end{document}